\documentstyle[preprint,prl,aps]{revtex}

\begin{document}

\preprint{Submitted to {\it Physical Review Letters}}
\draft

\title{Noncommuting mixed states cannot be broadcast}

\author{Howard Barnum, Carlton M.~Caves, Christopher A.~Fuchs,\\
Richard Jozsa,\cite{JozsaAddress} and
Benjamin Schumacher\cite{SchumacherAddress}}
\address{Center for Advanced Studies, Department of Physics and Astronomy,\\
University of New Mexico,
Albuquerque, New Mexico 87131--1156}

\date{\today}

\maketitle

\begin{abstract}
We show that, given a general mixed state for a quantum system, there
are no physical means for {\it broadcasting\/} that state onto two
separate quantum systems, even when the state need only be reproduced
marginally on the separate systems.  This result generalizes and extends
the standard no-cloning theorem for pure states.
\end{abstract}


The fledgling field of quantum information theory\cite{Bennett95}
draws attention to fundamental questions about what is physically
possible and what is not.  An example is the
theorem\cite{Wootters82,Dieks82} that there are no physical means
by which an {\it unknown pure\/} quantum state can be reproduced
or copied---a result summarized by the phrase ``quantum
states cannot be cloned.''  In this paper we formulate and prove an
impossibility theorem that extends and generalizes the pure-state
no-cloning theorem to mixed quantum states.  The theorem answers the
question: are there any physical means for {\it broadcasting\/}
an unknown quantum state, pure or mixed, onto two separate
quantum systems?  By broadcasting we mean that the marginal density
operator of each of the separate systems is the same as the state to be
broadcast.

The pure-state ``no-cloning'' theorem\cite{Wootters82,Dieks82}
prohibits broadcasting pure states, for the only way to broadcast
a pure state $|\psi\rangle$ is to put the two systems in the product
state $|\psi\rangle\otimes|\psi\rangle$, i.e., to clone $|\psi\rangle$.
Things are more complicated when the states are mixed.  A mixed-state
no-cloning theorem is not sufficient to demonstrate no-broadcasting,
for there are many conceivable ways to broadcast a mixed state $\rho$
without the joint state being in the product form $\rho\otimes\rho$,
the mixed-state analog of cloning; the systems might be correlated
or entangled in such a way as to give the right marginal density
operators.  For instance, if the density operator has the spectral
decomposition $\rho=\sum_b\lambda_b|b\rangle\langle b|$, a potential
broadcasting state is the highly correlated joint state
$\tilde\rho=\sum_b\lambda_b|b\rangle|b\rangle\langle b|\langle b|$,
which, though not of the product form $\rho\otimes\rho$, reproduces
the correct marginal probability distributions.

The general problem, posed formally, is this.  A quantum system AB is
composed of two parts, A and B, each having an $N$-dimensional Hilbert
space.  System A is secretly prepared in one state from a set
${\cal A}\!=\!\{\rho_0,\rho_1\!\}$ of two quantum states.  System B,
slated to receive the unknown state, is in a standard
quantum state $\Sigma$.  The initial state of the composite system AB
is the product state $\rho_s\otimes\Sigma$, where $s=0$ or 1 specifies
which state is to be broadcast.  We ask whether there is any physical
process $\cal E$, consistent with the laws of quantum theory, that leads
to an evolution of the form
$\rho_s\otimes\Sigma\rightarrow{\cal E}(\rho_s\otimes\Sigma)=\tilde\rho_s$,
where $\tilde\rho_s$ is {\it any\/} state on the $N^2$-dimensional
Hilbert space AB such that
\begin{equation}
{\rm tr}_{\scriptscriptstyle {\rm A}}(\tilde\rho_s)=\rho_s
\mbox{\qquad\rm and\qquad}
{\rm tr}_{\scriptscriptstyle {\rm B}}(\tilde\rho_s)=\rho_s\;.
\label{require}
\end{equation}
Here ${\rm tr}_{\scriptscriptstyle {\rm A}}$ and
${\rm tr}_{\scriptscriptstyle {\rm B}}$ denote partial traces over A and
B.  If there is an ${\cal E}$ that satisfies Eq.~(\ref{require}) for both
$\rho_0$ and $\rho_1$, then the set ${\cal A}$ can be {\it broadcast}.
A special case of broadcasting is the evolution specified by
${\cal E}(\rho_s\otimes\Sigma)=\rho_s\otimes\rho_s$; we reserve the word
{\it cloning\/} for this strong form of broadcasting.

The most general action $\cal E$ on AB consistent with quantum theory
is to allow AB to interact unitarily with an auxiliary quantum system
C in some standard state and thereafter to ignore the auxiliary
system\cite{Kraus83}; that is,
\begin{equation}
{\cal E}(\rho_s\otimes\Sigma)=
{\rm tr}_{\scriptscriptstyle {\rm C}}\!\left(U(
\rho_s\otimes\Sigma\otimes\Upsilon) U^\dagger\right),
\label{process}
\end{equation}
for some auxiliary system C, some standard state $\Upsilon$ on C, and
some unitary operator $U$ on ABC.  We show that such an evolution can
lead to broadcasting if and only if $\rho_0$ and $\rho_1$ commute.  This
result strikes close to the heart of the difference between the
classical and quantum theories, because it provides another physical
distinction between {\it commuting\/} and {\it noncommuting\/} states.
We further show that $\cal A$ is clonable if and only if $\rho_0$ and
$\rho_1$ are identical or orthogonal ($\rho_0\rho_1=0$).

To see that the set $\cal A$ can be broadcast when the states commute,
we do not need to attach an auxiliary system.  Since orthogonal pure
states can be cloned, broadcasting can be obtained by cloning the
simultaneous eigenstates of $\rho_0$ and $\rho_1$.
Let $|b\rangle$, $b=1,\ldots ,N$, be an orthonormal basis for A in
which both $\rho_0$ and $\rho_1$ are diagonal, and let their spectral
decompositions be $\rho_s=\sum_b\lambda_{sb}|b\rangle\langle b|$.
Consider any unitary operator $U$ on AB consistent with
$U|b\rangle|1\rangle=|b\rangle|b\rangle$. If we choose
$\Sigma=|1\rangle\langle1|$ and let
\begin{equation}
\tilde\rho_s=
U(\rho_s\otimes\Sigma)U^\dagger=
\sum_b\lambda_{sb}|b\rangle|b\rangle\langle b|\langle b|\;,
\label{waldo}
\end{equation}
we immediately have that $\tilde\rho_0$ and $\tilde\rho_1$ satisfy
Eq.~(\ref{require}).

The converse of this statement---that if $\cal A$ can be broadcast,
$\rho_0$ and $\rho_1$ commute---is more difficult to prove.  Our
proof is couched in  terms of the concept of {\it fidelity\/} between
two density operators.  The fidelity $F(\rho_0,\rho_1)$ is defined by
\begin{equation}
F(\rho_0,\rho_1)={\rm tr}\sqrt{\rho_0^{1/2}\rho_1\rho_0^{1/2}\,}\;,
\label{fidel}
\end{equation}
where for any positive operator $O$, i.e., any Hermitian operator with {\it
nonnegative\/} eigenvalues, $O^{1/2}$ denotes its unique
positive square root. (Note that Ref.~\cite{Jozsa94a} defines fidelity
to be the square of the present quantity.)  Fidelity is an analogue of the
modulus of the inner product for pure states\cite{Uhlmann76,Jozsa94a}
and can be interpreted as a measure of distinguishability for quantum
states: it ranges between 0 and 1, reaching O if and only
if the states are orthogonal and reaching 1 if and only
if $\rho_0=\rho_1$.  It is  invariant under the interchange
$0\leftrightarrow1$ and under the transformation
$\rho_0\rightarrow U\rho_0 U^\dagger$,
$\rho_1\rightarrow U\rho_1 U^\dagger$ for any unitary operator
$U$\cite{Jozsa94a,Fuchs95b}.  Also, from the properties of the direct
product, one has that $F(\rho_0\otimes\sigma_0,\rho_1\otimes\sigma_1)=
F(\rho_0,\rho_1)F(\sigma_0,\sigma_1)$.

Another reason $F(\rho_0,\rho_1)$ defines a good notion of
distinguishability \cite{Wootters81} is that it equals the minimal
overlap between the probability distributions $p_0(b)={\rm tr}(\rho_0 E_b)$
and $p_1(b)={\rm tr}(\rho_1 E_b)$ generated by a generalized
measurement or {\it positive operator-valued measure\/} (POVM)
$\{E_b\}$\cite{Kraus83}. That is\cite{Fuchs95b},
\begin{equation}
F(\rho_0,\rho_1)=\min_{\{E_b\}}\sum_b\sqrt{{\rm tr}(\rho_0 E_b)}
\sqrt{{\rm tr}(\rho_1 E_b)}\;,
\label{uncle}
\end{equation}
where the minimum is taken over all sets of positive operators $\{E_b\}$
such that $\sum_b E_b=\openone$.  This representation of fidelity
has the advantage of being defined operationally in terms of measurements.
We call a POVM that achieves the minimum in Eq.~(\ref{uncle}) an
{\it optimal\/} POVM.

One way to see the equivalence of Eqs.~(\ref{uncle}) and (\ref{fidel})
is through the Schwarz inequality for the operator inner product
${\rm tr}(AB^\dagger)$:
${\rm tr}(AA^\dagger)\,{\rm tr}(BB^\dagger)\ge|{\rm tr}(AB^\dagger)|^2$,
with equality if and only if $A=\alpha B$ for some constant $\alpha$.
Going through this exercise is useful because it leads directly to the
proof of the no-broadcasting theorem.  Let $\{E_b\}$ be any POVM and
let $U$ be any unitary operator.  Using the cyclic property of the trace
and the Schwarz inequality, we have that
\begin{eqnarray}
\sum_b
&\mbox{}&
\sqrt{{\rm tr}(\rho_0 E_b)}\sqrt{{\rm tr}(\rho_1 E_b)}
\nonumber\\
&\mbox{}&
=\sum_b\sqrt{{\rm tr}\!
\left(U\rho_0^{1/2}E_b\,\rho_0^{1/2}U^\dagger\right)}
\sqrt{{\rm tr}\!\left(\rho_1^{1/2}E_b\,\rho_1^{1/2}\right)}
\nonumber\\
&\mbox{}&
\ge\sum_b\left|{\rm tr}\!
\left(U\rho_0^{1/2}E_b^{1/2}E_b^{1/2}\rho_1^{1/2}\right)\right|
\eqnum{I}\\
&\mbox{}&
\ge\left|\sum_b{\rm tr}\!
\left(U\rho_0^{1/2}E_b\rho_1^{1/2}\right)\right|
=\Bigl|\,{\rm tr}\!\left(U\rho_0^{1/2}\rho_1^{1/2}\right)\Bigr|\;.
\label{Quebecois}
\end{eqnarray}
We can use the freedom in $U$ to make the inequality as tight as
possible.  To do this, we recall\cite{Jozsa94a,Schatten60} that
$\max|{\rm tr}(V\!O)|={\rm tr}\sqrt{O^\dagger O}$, where $O$ is any
operator and the maximum is taken over all unitary operators
$V$.  The maximum is achieved only by those $V$ such that
$V\!O=\sqrt{O^\dagger O}$; that there exists at least one such $V$
is insured by the operator polar decomposition theorem\cite{Schatten60}.
Therefore, by choosing
\begin{equation}
U\rho_0^{1/2}\rho_1^{1/2}=\sqrt{\rho_1^{1/2}\rho_0\rho_1^{1/2}\,}\;,
\label{rickets}
\end{equation}
we get that
$\sum_b\!\sqrt{{\rm tr}(\rho_0 E_b)}
\sqrt{{\rm tr}(\rho_1 E_b)}\ge F(\rho_0,\rho_1)$.

To find optimal POVMs, we consult the conditions for equality in
Eq.~(\ref{Quebecois}).  These arise from step I and the one following
it: a POVM is optimal if and only if
\begin{equation}
U\rho_0^{1/2}E_b^{1/2}=\mu_b\rho_1^{1/2}E_b^{1/2}
\label{scurvy}
\end{equation}
and
\begin{equation}
{\rm tr}\!\left(U\rho_0^{1/2}E_b\rho_1^{1/2}\right)=
\mu_b\,{\rm tr}(\rho_1 E_b)\ge0\;\Leftrightarrow\;\mu_b\ge0\;.
\label{mangoroots}
\end{equation}
When $\rho_1$ is invertible, Eq.~(\ref{scurvy}) becomes
\begin{equation}
M E_b^{1/2}=\mu_b E_b^{1/2}\;,
\label{Fontaine}
\end{equation}
where
\begin{equation}
M=\rho_1^{-1/2}U\rho_0^{1/2}=
\rho_1^{-1/2}\sqrt{\rho_1^{1/2}\rho_0\rho_1^{1/2}\,}\rho_1^{-1/2}
\label{gop}
\end{equation}
is a positive operator. Therefore one way to satisfy Eq.~(\ref{scurvy})
with $\mu_b\ge0$ is to take $E_b=|b\rangle\langle b|$,
where the vectors $|b\rangle$ are an orthonormal eigenbasis for $M$,
with $\mu_b$ chosen to be the eigenvalue of $|b\rangle$.  When $\rho_1$
is noninvertible, there are still optimal POVMs.  One can choose the
first $E_b$ to be the projector onto the null subspace of $\rho_1$; in
the support of $\rho_1$, i.e., the orthocomplement of the null subspace,
$\rho_1$ is invertible, so one can construct the analogue of $M$ and
proceed as for an invertible $\rho_1$.  Note that if both $\rho_0$ and
$\rho_1$ are in\-vert\-ible, $M$ is invertible.

We begin the proof of the no-broadcasting theorem by using Eq.~(\ref{uncle})
to show that fidelity cannot decrease under the operation of partial
trace; this gives rise to an elementary constraint on all potential
broadcasting processes $\cal E$.  Suppose Eq.~(\ref{require}) is
satisfied for the process $\cal E$ of Eq.~(\ref{process}), and let
$\{E_b\}$ denote an optimal POVM for distinguishing $\rho_0$
and $\rho_1$.  Then, for each $s$,
${\rm tr}\bigl(\tilde\rho_s(E_b\otimes\openone)\bigr)=
{\rm tr}_{\scriptscriptstyle{\rm A}}\bigl(
{\rm tr}_{\scriptscriptstyle{\rm B}}(\tilde\rho_s)E_b\bigr)=
{\rm tr}_{\scriptscriptstyle{\rm A}}(\rho_s E_b)$;
it follows that
\begin{eqnarray}
F_{\scriptscriptstyle{\rm A}}(\rho_0,\rho_1)
&\equiv&
\sum_b\sqrt{{\rm tr}\bigl(\tilde\rho_0(E_b\otimes\openone)\bigr)}
\sqrt{{\rm tr}\bigl(\tilde\rho_1(E_b\otimes\openone)\bigr)}
\nonumber\\
&\ge&
\min_{\{\tilde E_c\}}\,\sum_c\sqrt{{\rm tr}(\tilde\rho_0\tilde E_c)}
\sqrt{{\rm tr}(\tilde\rho_1\tilde E_c)}
\nonumber\\
&=&
F(\tilde\rho_0,\tilde\rho_1)
\;.
\label{gaggle}
\end{eqnarray}
Here $F_{\scriptscriptstyle{\rm A}}(\rho_0,\rho_1)$ denotes the fidelity
$F(\rho_0,\rho_1)$; the subscript A emphasizes that
$F_{\scriptscriptstyle{\rm A}}(\rho_0,\rho_1)$ stands for the
particular representation on the first line.  The inequality in
Eq.~(\ref{gaggle}) comes from the fact that $\{E_b\otimes\openone\}$
might not be an optimal POVM for distinguishing $\tilde\rho_0$ and
$\tilde\rho_1$; this demonstrates the said partial trace property.  Similarly
it follows that
\begin{eqnarray}
F_{\scriptscriptstyle {\rm B}}(\rho_0,\rho_1)
&\equiv&
\sum_b\sqrt{{\rm tr}\bigl(\tilde\rho_0(\openone\otimes E_b)\bigr)}
\sqrt{{\rm tr}\bigl(\tilde\rho_1(\openone\otimes E_b)\bigr)}
\nonumber\\
&\ge&
F(\tilde\rho_0,\tilde\rho_1)\;,
\label{jojo}
\end{eqnarray}
where the subscript B emphasizes that
$F_{\scriptscriptstyle {\rm B}}(\rho_0,\rho_1)$ stands for the
representation on the first line.

On the other hand, we can just as easily derive an inequality that is opposite
to Eqs.~(\ref{gaggle}) and (\ref{jojo}).  By the direct product formula and
the invariance of fidelity under unitary transformations,
\begin{eqnarray}
F(\rho_0,\rho_1)
&=&
F(\rho_0\otimes\Sigma\otimes\Upsilon,\rho_1\otimes\Sigma\otimes\Upsilon)
\\
&=&
F\Bigl(U(\rho_0\otimes\Sigma\otimes\Upsilon)U^\dagger,
U(\rho_1\otimes\Sigma\otimes\Upsilon)U^\dagger\Bigr)\;.
\rule{0mm}{5mm}\nonumber
\end{eqnarray}
Therefore, by the partial-trace property,
\begin{eqnarray}
&\mbox{}&F(\rho_0,\rho_1)\\
&\mbox{}&\le
F\!\Bigl({\rm tr}_{\scriptscriptstyle {\rm C}}\!
\left(U(\rho_0\otimes\Sigma\otimes\Upsilon)U^\dagger\right),
{\rm tr}_{\scriptscriptstyle {\rm C}}\!
\left(U(\rho_1\otimes\Sigma\otimes\Upsilon)U^\dagger\right)\!\Bigr),
\rule{0mm}{5mm}
\nonumber
\end{eqnarray}
or, more succinctly,
\begin{equation}
F(\rho_0,\rho_1)\le
F\Bigl({\cal E}(\rho_0\otimes\Sigma),
{\cal E}(\rho_1\otimes\Sigma)\Bigr)=
F(\tilde\rho_0,\tilde\rho_1)\;.
\label{Wilma}
\end{equation}

The elementary constraint now follows, for the only way to
maintain Eqs.~(\ref{gaggle}), (\ref{jojo}), {\it and\/}
(\ref{Wilma}) is with strict equality.  In other words,
we have that if the set $\cal A$ can be broadcast, then there are
density operators $\tilde\rho_0$ and $\tilde\rho_1$ on AB satisfying
Eq.~(\ref{require}) {\it and}
\begin{equation}
F_{\scriptscriptstyle{\rm A}}(\rho_0,\rho_1)=
F(\tilde\rho_0,\tilde\rho_1)=F_{\scriptscriptstyle{\rm B}}(\rho_0,\rho_1)\;.
\label{SmootGibson}
\end{equation}

Let us pause at this point to consider the restricted question of
cloning. If $\cal A$ is to be clonable, there must exist a process
$\cal E$ such that $\tilde\rho_s=\rho_s\otimes\rho_s$ for $s=0,1$.
But then, by Eq.~(\ref{SmootGibson}), we must have
\begin{equation}
F(\rho_0,\rho_1)=F(\rho_0\otimes\rho_0,\rho_1\otimes\rho_1)=
F(\rho_0,\rho_1)^2,
\label{Spanky}
\end{equation}
which means that $F(\rho_0,\rho_1)=1$ or 0, i.e., $\rho_0$ and $\rho_1$
are identical or orthogonal.  There can be no cloning for density operators
with nontrivial fidelity.  The converse, that orthogonal and identical
density operators can be cloned, follows, in the first case, from the fact
that they can be distinguished by measurement and, in the second case,
because they need not be distinguished at all.

Like the pure-state no-cloning theorem\cite{Wootters82,Dieks82}, this
no-cloning result for mixed states is a consistency requirement for
the axiom that quantum measurements cannot distinguish nonorthogonal
states with perfect reliability.  If nonorthogonal quantum states
could be cloned, there would exist a measurement procedure for
distinguishing those states with arbitrarily high reliability: one
could make measurements on enough copies of the quantum state to
make the probability of a correct inference of its identity
arbitrarily high.  That this consistency requirement, as expressed
in Eq.~(\ref{SmootGibson}), should also exclude more general kinds
of broadcasting problems is not immediately obvious.  Nevertheless,
this is the content of our claim that Eq.~(\ref{SmootGibson}) generally
cannot be satisfied; any broadcasting process can be viewed as
creating distinguishability {\it ex nihilo\/} with respect to
measurements on the larger Hilbert space AB.  Only for the case
of commuting density operators does broadcasting not create
any extra distinguishability.

We now show that Eq.~(\ref{SmootGibson}) implies that $\rho_0$ and
$\rho_1$ commute.  To simplify the exposition, we assume that
$\rho_0$ and $\rho_1$ are invertible.  We proceed by studying the
conditions necessary for the representations
$F_{\scriptscriptstyle{\rm A}}(\rho_0,\rho_1)$ and
$F_{\scriptscriptstyle{\rm B}}(\rho_0,\rho_1)$ in
Eqs.~(\ref{gaggle}) and (\ref{jojo}) to equal
$F(\tilde\rho_0,\tilde\rho_1)$.  Recall that the optimal POVM
$\{E_b\}$ for distinguishing $\rho_0$ and $\rho_1$ can be chosen
so that the POVM elements $E_b=|b\rangle\langle b|$ are a complete
set of orthogonal one-dimensional projectors onto orthonormal
eigenstates of $M$.  Then, repeating the steps leading from
Eqs.~(\ref{Quebecois}) to (\ref{mangoroots}), one finds that
the necessary conditions for equality in Eq.~(\ref{SmootGibson})
are that each $E_b\otimes\openone=(E_b\otimes\openone)^{1/2}$ and
each $\openone\otimes E_b=(\openone\otimes E_b)^{1/2}$ satisfy
\begin{eqnarray}
\tilde U\tilde\rho_0^{1/2}(\openone\otimes E_b)&=&
\alpha_b\,\tilde\rho_1^{1/2}(\openone\otimes E_b)\;,
\label{parceltree}
\\
\tilde V\tilde\rho_0^{1/2}(E_b\otimes\openone)&=&
\beta_b\,\tilde\rho_1^{1/2}(E_b\otimes\openone)\;,
\rule{0mm}{5mm}
\label{skypost}
\end{eqnarray}
where $\alpha_b$ and $\beta_b$ are nonnegative numbers and $\tilde U$
and $\tilde V$ are unitary operators satisfying
\begin{equation}
\tilde U\tilde\rho_0^{1/2}\tilde\rho_1^{1/2}=
\tilde V\tilde\rho_0^{1/2}\tilde\rho_1^{1/2}=
\sqrt{\tilde\rho_1^{1/2}\tilde\rho_0\tilde\rho_1^{1/2}\,}\;.
\label{temptation}
\end{equation}
Although $\rho_0$ and $\rho_1$ are assumed invertible, one cannot
demand that $\tilde\rho_0$ and $\tilde\rho_1$ be invertible---a glance
at Eq.~(\ref{waldo}) shows that to be too restrictive.  This means
that $\tilde U$ and $\tilde V$ need not be the same.  Also we cannot
assume that there is any relation between $\alpha_b$ and $\beta_b$.

The remainder of the proof consists in showing that Eqs.~(\ref{parceltree})
through (\ref{temptation}), which are necessary (though perhaps not
sufficient) for broadcasting, are nevertheless restrictive enough to imply
that $\rho_0$ and $\rho_1$ commute.  The first step is to sum over $b$ in
Eqs.~(\ref{parceltree}) and (\ref{skypost}).  Defining the positive
operators
\begin{equation}
G=\sum_b\alpha_b|b\rangle\langle b|
\mbox{\qquad and\qquad}
H=\sum_b\beta_b|b\rangle\langle b|\;,
\end{equation}
we obtain
\begin{equation}
\tilde U\tilde\rho_0^{1/2}=\tilde\rho_1^{1/2}(\openone\otimes G)
\mbox{\quad and\quad}
\tilde V\tilde\rho_0^{1/2}=\tilde\rho_1^{1/2}(H\otimes\openone)\;.
\label{vituperative}
\end{equation}

The next step is to demonstrate that $G$ and $H$ are invertible and,
in fact, equal to each other.  Multiplying the two equations in
Eq.~(\ref{vituperative}) from the left by
$\tilde\rho_0^{1/2}\tilde U^\dagger$ and
$\tilde\rho_0^{1/2}\tilde V^\dagger$, respectively, and partial tracing
the first over A and the second over B, we get
\begin{equation}
\rho_0={\rm tr}_{\scriptscriptstyle{\rm A}}
\Bigl(\tilde\rho_0^{1/2}\tilde U^\dagger\tilde\rho_1^{1/2}\Bigr)G
\;\;\mbox{and}\;\;
\rho_0={\rm tr}_{\scriptscriptstyle{\rm B}}
\Bigl(\tilde\rho_0^{1/2}\tilde V^\dagger\tilde\rho_1^{1/2}\Bigr)H\;.
\label{Lewis}
\end{equation}
Since, by assumption, $\rho_0$ is invertible, it follows that
$G$ and $H$ are invertible.  Returning to Eq.~(\ref{vituperative}),
multiplying both parts from the left by $\tilde \rho_1^{1/2}$ and
tracing over A and B, respectively, we obtain
\begin{equation}
{\rm tr}_{\scriptscriptstyle{\rm A}}
\Bigl(\tilde\rho_1^{1/2}\tilde U\tilde\rho_0^{1/2}\Bigr)=\rho_1G
\mbox{\quad and\quad}
{\rm tr}_{\scriptscriptstyle{\rm B}}
\Bigl(\tilde\rho_1^{1/2}\tilde V\tilde\rho_0^{1/2}\Bigr)=\rho_1H\;.
\label{miasma}
\end{equation}
Conjugating the two parts of Eq.~(\ref{miasma}) and inserting
the results into the two parts of Eq.~(\ref{Lewis}) yields
\begin{equation}
\rho_0=G\rho_1G\mbox{\qquad and\qquad}\rho_0=H\rho_1H\;.
\label{yellowbelly}
\end{equation}
This shows that $\mbox{$G=H$}$, because these equations have a
unique positive solution, namely the operator $M$ of Eq.~(\ref{gop}).
This can be seen by multiplying Eq.~(\ref{yellowbelly})
from the left and right by $\rho_1^{1/2}$ to get
$
\rho_1^{1/2}\rho_0\rho_1^{1/2}=\bigl(\rho_1^{1/2}G\rho_1^{1/2}\bigr)^{\!2}
$.
The positive operator $\rho_1^{1/2}G\rho_1^{1/2}$ is thus
the unique positive square root of $\rho_1^{1/2}\rho_0\rho_1^{1/2}$.

Knowing that $\mbox{$G=H=M$}$, we return to Eq.~(\ref{vituperative}).  The two
parts, taken together, imply that
\begin{equation}
\tilde V^\dagger\tilde U\tilde\rho_0^{1/2}=\tilde\rho_0^{1/2}
(M^{-1}\!\otimes M)\;.
\label{BiancaJ}
\end{equation}
If $|b\rangle$ and $|c\rangle$ are eigenvectors of $M$, with eigenvalues
$\mu_b$ and $\mu_c$, Eq.~(\ref{BiancaJ}) implies that
\begin{equation}
\tilde V^\dagger\tilde U\Bigl(\tilde\rho_0^{1/2}|b\rangle|c\rangle\Bigr)=
\frac{\mu_c}{\mu_b}\Bigl(\tilde\rho_0^{1/2}|b\rangle|c\rangle\Bigr)\;.
\label{salsa}
\end{equation}
This means that $\tilde\rho_0^{1/2}|b\rangle|c\rangle$
is zero or it is an eigenvector of the unitary operator
$\tilde V^\dagger\tilde U$.  In the latter case, since the eigenvalues
of a unitary operator have modulus 1, it must be true that
$\mu_b=\mu_c$.  Hence we can conclude that
\begin{equation}
\tilde\rho_0^{1/2}|b\rangle|c\rangle=0
\mbox{\qquad when\qquad}\mu_b\ne\mu_c\;.
\label{Benjamin}
\end{equation}
This is enough to show that $M$ and $\rho_0$ commute and hence
$[\rho_0,\rho_1]=0$.  Consider the matrix element
\begin{eqnarray}
\langle b'|(M\rho_0-\rho_0M)|b\rangle
&=&
(\mu_{b'}-\mu_b)\langle b'|\rho_0|b\rangle\nonumber\\
&=&
(\mu_{b'}-\mu_b)\sum_c
\langle b'|\langle c|\,\tilde\rho_0|c\rangle|b\rangle\;.
\end{eqnarray}
If $\mu_b=\mu_{b'}$, this is automatically zero.  If, on the other hand,
$\mu_b\ne\mu_{b'}$, then the sum over $c$ must vanish by Eq.~(\ref{Benjamin}).
It follows that $\rho_0$ and $M$ commute.  Hence, using
Eq.~(\ref{yellowbelly}),
\begin{equation}
\rho_1\rho_0=M^{-1}\rho_0M^{-1}\rho_0=
\rho_0M^{-1}\rho_0M^{-1}=\rho_0\rho_1\;.
\end{equation}
This completes the proof that noncommuting quantum states cannot be
broadcast.

Note that, by the same method as above,
$\tilde\rho_1^{1/2}|b\rangle|c\rangle=0$ when $\mu_b\ne\mu_c$.  This
condition, along with Eq.~(\ref{Benjamin}), determines the conceivable
broadcasting states, in which the correlations between the systems A
and B range from purely classical to purely quantum.  For example,
since $\rho_0$ and $\rho_1$ commute, the states of Eq.~(\ref{waldo}) satisfy
these conditions, but so do the perfectly entangled pure states
$\sum_b\sqrt{\lambda_{sb}}|b\rangle|b\rangle$.  Not all such broadcasting
states can be realized by a physical process $\cal E$, but sufficient
conditions for realizability are not known.

In closing, we mention an application of this result.  In
some versions of quantum cryptography\cite{Bennett92b}, the legitimate
users of a communication channel encode the bits 0 and 1 into
nonorthogonal pure states.  This is done to ensure that any
eavesdropping is detectable, since
eavesdropping necessarily disturbs the states sent
to the legitimate receiver\cite{Bennett92a}.  If
the channel is noisy, however, causing the bits to evolve to
noncommuting mixed states, the detectability of eavesdropping is
no longer a given.  The result presented here shows that
there are no means available for an eavesdropper to obtain the
signal, noise and all, intended for the legitimate receiver without
in some way changing the states sent to the receiver.

We thank Richard Hughes for useful discussions.  This work was supported
in part by the Office of Naval Research (Grant No.~N00014-93-1-0116).


\end{document}